# Relating the optical absorption coefficient of nanosheet dispersions to the intrinsic monolayer absorption

Keith R. Paton and Jonathan N Coleman*

*School of Physics and CRANN & AMBER Research Centres, Trinity College Dublin, Dublin 2, Ireland*

*colemaj@tcd.ie

Abstract: The concentration of nanosheet suspensions is an important technological parameter which is commonly measured by optical spectroscopy using the absorption coefficient to transform absorbance into concentration. However for all 2D materials, the absorption coefficient is poorly known, resulting in potentially large errors in measured concentration. Here we derive an expression relating the optical absorption coefficient of an isotropic ensemble of nanosheets to the intrinsic monolayer absorption. This has allowed us to calculate the absorption coefficients for suspensions of graphene, $MoS_2$ and other 2D materials and allows estimation of the monolayer absorption for new materials from careful measurement of the suspension absorption coefficient.

Introduction

Liquid phase exfoliation of layered materials has become one of the most widely used methods to obtain 2D nanosheets in an easily processable form [1-8]. Nanosheet suspensions (a.k.a. dispersions or inks) have been shown to be ideally suited to production of printed electronics [9,10], including devices such as LEDs, battery and supercapacitor electrodes [11-16], photo-detectors [17] and hydrogen evolution catalysis [18-22], as well as additives in composites [23-26]. While the production of these dispersions can be scaled to industrial levels[8], methods to reliably and rapidly characterise the material produced are limited. Measurement of the concentration of nanosheets in the dispersion is widely achieved by optical absorption spectroscopy, through the application of the Beer-Lambert Law. While this has been widely used for a range of 2-D nanosheet dispersions, it requires an accurate value for the absorption coefficient (or more practically the extinction coefficient[27]). This in turn is usually obtained by removing the liquid from the dispersion, either by filtering or evaporation, and weighing the resulting solid (taking account of any surfactant or solvent residues). Despite the simplicity of the procedure to obtain this important parameter, the values reported in the literature vary widely, from as low as 1043 ml·mg$^{-1}$·m$^{-1}$ to as high as 6600 ml·mg$^{-1}$·m$^{-1}$ for graphene [28, 29]. Similar variations exist for other 2D materials such as MoS$_2$ (see SI). It is not clear which papers are correct as theoretical values of the suspension absorption coefficient are not available.

However, this is a problem which should be easily addressed. For a number of 2D materials, notably graphene, the amount of incident light absorbed by a single monolayer is known experimentally. This intrinsic material property is the primary factor controlling the absorption coefficient of a dispersion of nanosheets. Once it is known, it should be straightforward to derive a relationship between these quantities. However, to date such a calculation has not been published. In the present paper, we derive an expression relating the absorption coefficient for dispersions of 2-D nanosheets to the absorption of a monolayer. This relationship will allow the calculation of dispersion absorption coefficients from theoretical estimates of monolayer absorption and will allow the validation of experimental values of absorption coefficient. In addition, it will allow the estimation of monolayer absorption from careful measurements of absorption coefficient.

Optical spectrometers physically measure the transmission of light, T, (defined as the ratio of transmitted, $I$, to incident, $I_0$, light intensity). However, the data is often outputted as the absorbance, which we will refer to here as $A_T$. This parameter is generally defined as $A_T = -\log_{10} T$ and it is this quantity which is automatically outputted by the spectrometer software (i.e. not $-\ln T$). The absorbance is useful because it is directly proportional to the quantity of absorbing material: $A_T = \alpha C L$ where C is the concentration, defined as the dispersed mass/dispersion volume. Here the proportionality constant, $\alpha$, is the absorption coefficient which tends to be poorly known (L is the cell length, the distance the beam travels through the vessel containing the liquid).

The simplest way to calculate $\alpha$ is via the absorption, A, which is the fractional light intensity change as the beam travels through the sample: $A = (T_0 - T)/T_0$. Neglecting reflections, we have $A = 1 - T$. For a dilute solution, where $\alpha C L$ is small, it is straightforward to show that $A = \alpha C L / \log_{10} e$, where $e$=2.72. Thus, calculation of the absorbance will allow us to find the absorption coefficient.

The absorption of a dispersion of nanosheets is just the sum of the absorptions of all individual nanosheets. To calculate this we must consider that at any given instant, the nanosheets are randomly distributed throughout the liquid with isotropic orientation distribution. To calculate the total absorption, we consider a nanosheet whose orientation is defined by the polar angle, $\theta$, and azimuthal angle, $\phi$ associated with the unit vector normal to its basal plane, $\hat{\mathbf{n}}$ (see Figure 1A). The contribution to the absorption from all nanosheets with this orientation is given by

$$dA = A_{NS}(\theta,\phi) N_{\Omega} d\Omega \tag{1}$$

where $A_{NS}(\theta,\phi)$ is the absorption of a single nanosheet of this orientation, $N_{\Omega}$ is the number of nanosheets per unit solid angle and $d\Omega$ is the differential solid angle defined by $\theta$ and $\phi$ and is given by $d\Omega = \sin\theta d\theta d\phi$ (Figure 1B). $N_{\Omega}$ is just the total number of nanosheets multiplied by the nanosheet orientation distribution function, $\Gamma_{NS}$. The latter parameter is the fraction of nanosheets per unit solid angle and for an isotropic distribution is given by $\Gamma_{NS} = 1/2\pi$ (see SI). This allows us to write $N_{\Omega} = N_V \Lambda_{Beam} L / 2\pi$ where $N_V$ is the number of nanosheets per unit volume, $\Lambda_{Beam}$ is the area of the beam in the x-y plane and L is the cell length. This parameter can be written in terms of the nanosheet concentration:

$$N_{\Omega} = \frac{C \Lambda_{Beam} L}{2\pi \rho_{NS} \Lambda_{NS} t_{NT}} \tag{2}$$

where $\rho_{NS}$, $\Lambda_{NS}$ and $t_{NS}$ are the nanosheet density, area and thickness respectively.

This allows us to write the total absorption of the dispersion as

$$A = \frac{C \Lambda_{Beam} L}{2\pi \rho_{NS} \Lambda_{NS} t_{NT}} \int_{\theta=0}^{\pi} \int_{\phi=0}^{\pi} A_{NS}(\theta,\phi) \sin\theta d\theta d\phi \tag{3}$$

where the upper limits of integration are π because of the planar symmetry of the nanosheet, with absorption equivalent regardless of the direction of light propagation through the basal plane.

To calculate $A_{NS}(\theta,\phi)$, requires the realisation that nanosheets of different orientations absorb light differently for two reasons. Firstly, the projected area which the nanosheet presents to the beam depends on nanosheet orientation and secondly, the amount of light the nanosheet absorbs depends on the square of the cosine of the angle between the nanosheet basal plane and the electric field vector of the light (essentially Malus' law, see SI).

The fraction of the total light intensity absorbed by a single nanosheet is given by the fraction of beam area occluded by the nanosheet, $F_\Lambda$, times the fraction of light intensity incident on the nanosheet which is absorbed. This second parameter is given by $A_\parallel \cos^2\gamma$, where $A_\parallel$ is the intrinsic nanosheet absorption (i.e. when the electric field of the light is parallel to the nanosheet basal plane) and $\gamma$ is the angle between the electric field vector and the basal plane of the nanosheet. Combining these gives:

$$A_{NS}(\theta,\phi) = F_\Lambda(\theta,\phi) A_\parallel \cos^2\gamma(\theta,\phi) \tag{4}$$

Below, we will address $F_\Lambda(\theta,\phi)$, $A_\parallel$ and $\gamma(\theta,\phi)$ separately.

The parameter $F_\Lambda(\theta,\phi)$ represents the fraction of total beam area occluded by a nanosheet whose orientation is described by $\theta$ and $\phi$. This is simply the projection of the nanosheet area onto the plane perpendicular to the propagation direction of the light (here the x-y plane) and is given by

$$F_\Lambda(\theta,\phi) = \frac{\Lambda_{NS}}{\Lambda_{Beam}} \cos\Psi \tag{5}$$

where $\Psi$ is the angle between the basal plane of the nanosheet and the x-y plane. It can be shown (see SI) that $\cos\Psi = \sin\theta\sin\phi$, giving

$$F_\Lambda(\theta,\phi) = \frac{\Lambda_{NS}}{\Lambda_{Beam}} \sin\theta\sin\phi \tag{6}$$

The intrinsic absorption of the nanosheet is represented by $A_\parallel$ which describes the fractional reduction of intensity for light incident on the nanosheet. For a thin nanosheet comprised of N layers, this is approximately (see SI) given by

$$A_\parallel = NA_{ML} \tag{7}$$

where $A_{ML}$ is the intrinsic absorption of a monolayer (when the electric field vector is in the plane of the nanosheet). This parameter is defined by the electronic properties of the monolayer and is known for a number of materials (see below), for example $A_{ML} \approx 2.3\%$ for graphene at wavelengths >400 nm.[30]

The amount of light absorbed by the nanosheet depends on the angle between the plane of the nanosheet and the electric field vector, $\gamma(\theta,\phi)$. This depends on both the orientation of the nanosheet (described by $\theta$ and $\phi$) and by the direction of the electric field vector which we take as being in the x-y plane at an angle of β to the x-axis (see Figure 1C). Then, it can be shown (see SI) that

$$\cos^2\gamma(\theta,\phi) = 1 - \cos^2\theta\cos^2\beta - \sin^2\theta\cos^2\phi\sin^2\beta - 2\cos\theta\sin\theta\cos\phi\cos\beta\sin\beta \quad (8)$$

We can combine these expressions to given an equation for the absorption coefficient of the nanosheet dispersion:

$$\alpha = \frac{\log_{10} e}{2\pi\rho_{NS}d_0} A_{ML} \int_{\theta=0}^{\pi}\int_{\phi=0}^{\pi} \cos^2\gamma(\theta,\phi)\sin\theta\sin\phi\sin\theta d\theta d\phi \quad (9)$$

Here, we have used $t_{NS} = Nd_0$, where $d_0$ is the monolayer thickness and for brevity have not substituted $\cos^2\gamma$ from equation 8. Performing the integration reduces this equation to

$$\alpha(\lambda) = \frac{3\log_{10} e}{8\rho_{NS}d_0} A_{ML}(\lambda) \quad (10)$$

In this expression, we explicitly note that both $\alpha(\lambda)$ and $A_{ML}$ are usually functions of wavelength. We note that equation 10 is not a function of β, as expected for an isotropic nanosheet distribution. This shows that the absorption coefficient is polarisation independent.

This expression is extremely useful as it relates the measured absorption coefficient (α) to the intrinsic nanosheet absorption ($A_{ML}$). This can be used to either predict the absorption coefficient once $A_{ML}$ is known or to estimate $A_{ML}$ from measured optical spectra.

It is simplest to predict the absorption coefficient for graphene. As mentioned above, it has been shown that for graphene, $A_{ML,Gra}$=0.023 for wavelengths between ~400 nm and ~800 nm. Substituting this value into equation 10 gives a predicted value of $\alpha_{Gra}$=4237 ml.mg$^{-1}$m$^{-1}$, which is within the range of experimental values reported (see figure 2). Very recently, we carefully measured the absorption coefficient of graphene dispersions finding $\alpha_{750}$=4,861 Lg$^{-1}$m$^{-1}$, in very good agreement with the theoretical value.[31] For inorganic nanosheets, such as MoS$_2$ and WS$_2$, the nanosheet absorption is strongly dependent on wavelength. Further complicating the analysis, the absorption coefficient is often found to vary with nanosheet length,[21, 27] due to edge effects.[27] Therefore, equation 10 is only strictly valid at the specific wavelength where $\alpha_{Edge} = \alpha_{Basal}$. For MoS$_2$ it have been shown that this occurs at ~345 nm, which is a local minimum in the absorption spectrum. Monolayer MoS$_2$ has been measured as having an absorption of $A_{ML}$=0.07 at this local minimum (measured at 336 nm)[27]. We can therefore calculate a predicted value of $\alpha_{MoS2}(336nm)$=3631 ml.mg$^{-1}$m$^{-1}$ for dispersion of MoS$_2$ nansheets.

However, to compare absorption coefficients predicted by equation 10 to literature values is not straightforward. This is because the equation relating absorbance to transmission given above ($A_T = -\log_{10} T$), although widely used, is not strictly correct for dispersed nanoparticles. This is because, in an optical spectrometer, light is lost from the beam via both absorption and scattering. While the scattering component is generally significantly smaller than the absorption contribution, for nanomaterials it is not negligible. In reality the measured transmittance is controlled by the extinction coefficient, ε: $-\log_{10} T = \varepsilon CL$. The extinction coefficient includes contributions from both absorption and scattering such that $\varepsilon(\lambda) = \alpha(\lambda) + \sigma(\lambda)$ where σ is the scattering coefficient.[27] While these contributions can be separated using an integrating sphere to give the true value of α this is not widely done. Thus, the vast majority of papers which report values of "α" are actually reporting the extinction coefficient. This means most literature values purporting to be the absorption coefficient but actually giving the extinction coefficient should be larger than the theoretical value. Careful measurements of ε, α and σ for MoS$_2$ have shown that, at peak absorbance, α/ε~0.7-0.9. However, we note that this may not always be the case in practice because of the significant experimental errors associated with measuring the extinction coefficient.

Nevertheless, it is important to compare the predicted value of the absorption coefficient with the reported experimental values, even if these more accurately reflect the extinction coefficient. We have found a wide range of reported values for α for dispersions of several materials[1, 6, 8, 27-29, 32-48], and compared them with that predicted from equation 10. We have taken the value for $A_{ML}$ at the appropriate wavelength reported in reference [49] to calculate the predicted value. In figure 2 we have plotted the experimental value of α against $A_{ML}(\lambda)/\rho_{NS} d_0$ for these published values. Superimposed on this graph is the theoretical prediction (i.e. a straight line with a slope of $3\log_{10} e/8$). The data presented in figure 2 highlights the wide variation in experimental values for this important parameter, some reasons for which are outlined above. However, despite the spread, the values predicted from equation 10 do all fall within the range from experiments. For those materials where values of $A_{ML}$ are known, either measured or calculated, we are also able to predict values for the absorption coefficient for the dispersion. The predicted values using experimental values of $A_{ML}$ taken from ref [49] are shown in Table 1 below. A fuller list of published values of $A_{ML}$, including calculated values, and resulting predicted values for α is shown in the supporting information.

We can also use this result to estimate the monolayer absorption for a 2D material where this is not known. Using an integrating sphere, the absorption coefficient of GaS suspensions has been shown to be nanosheet-size-independent at 365 nm (very close to the bandedge, considerably below the peak absorption) with a value of $\alpha_{GaS} \approx 300$ ml.mg$^{-1}$m$^{-1}$. Taking the density as $\rho_{NS}$=3860 kg/m$^3$ and $d_0 \approx 0.5$ nm, we get *$A_{ML}(\lambda=365nm) \approx 0.0036$*. This value is rather low because the wavelength chosen is so close to the bandedge.

Finally, we note that when using equation 10 it must be kept in mind that it represents the absorption rather than the extinction coefficient. This limits how it can be used. The ideal approach would be to measure the absorption coefficient for the material under study using an integrating sphere and compare directly with the theoretical value. Failing that it would possible to use the fact that the extinction coefficient is generally 10-30% higher than the absorption coefficient in the resonant region.[27, 31] Applying this would allow the estimation of the extinction coefficient from the theoretical absorption coefficient. Such a procedure would allow estimations of nanosheet concentrations with acceptable accuracy.

In conclusion, we have derived an expression to calculate the value for the absorption coefficient for dispersions of 2D nanosheets, we have provided a benchmark to compare experimental values of the important characterisation parameter. While the optical absorption of monolayers has only been reported for a few of these materials, the expression can be applied to new materials as they are measured. By providing a theoretical value for the absorption coefficient, it is hoped that the current wide spread of experimental values will begin to narrow.

The research leading to these results has received funding from the European Union Seventh Framework Program under grant agreement n°604391 Graphene Flagship. We have also received support from the Science Foundation Ireland (SFI) funded centre AMBER (SFI/12/RC/2278). In addition, JNC acknowledges the European Research Council (SEMANTICS) and SFI (11/PI/1087) for financial support.

**Figures**

*Table 1: Predicted value of absorption coefficient, α, using equation 10 and references [30] and [49].*

| Material | $A_{ML}$ (%) | λ (nm) | α (ml mg$^{-1}$ m$^{-1}$) |
|---|---|---|---|
| Graphene | 2.3[30] | 400-800 | 4237 |
| MoS2 | 14.87[49] | 336 | 7719 |
| WS2 | 9.79[49] | 315 | 3429 |
| MoSe2 | 4.62[49] | 380 | 1661 |
| WSe2 | 10.05[49] | 383 | 2702 |

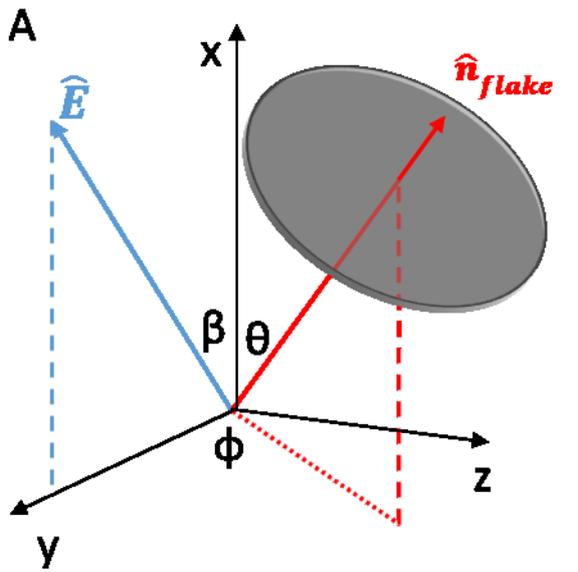

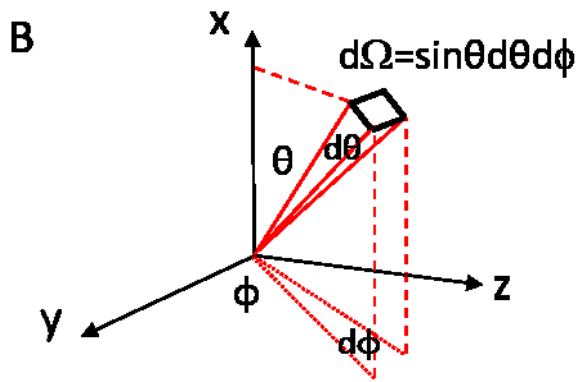

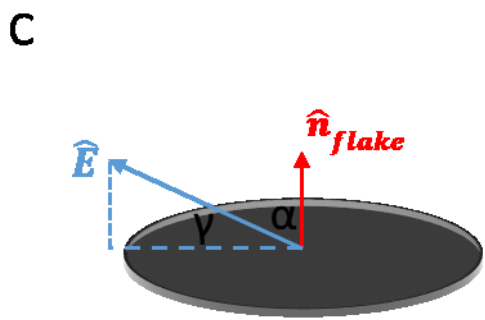

*Figure 1: A) Diagram showing geometry of flake orientation. Light is taken to propagate in the z-direction, with electric field vector making an angle β with the x-axis. The normal to the flake makes an angle θ with the x-axis, and ϕ with the y-axis. The flake is shown as offset from the origin for clarity. B) Schematic showing the construction of the differential solid angle dΩ. C) Diagram showing angle between the flake and electric field vector of the light.*

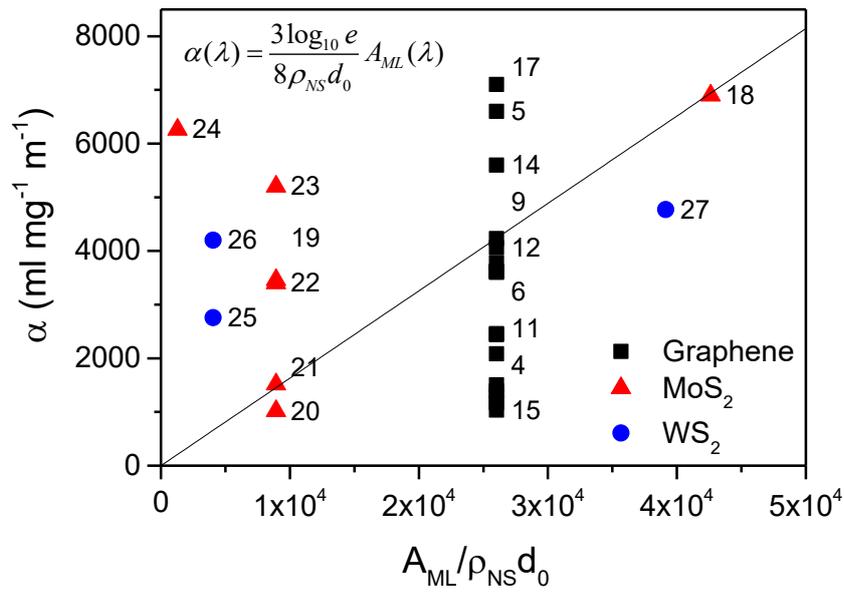

*Figure 2: Reported values of absorption coefficient for dispersions of graphene (black squares), MoS$_2$ (red triangles) and WS$_2$ (blue circles). These have been plotted against the calculated value of $A_{ML}/(\rho_{NS}d_0)$ with the value for $A_{ML}$ taken from ref [49] and [30] at the same wavelength as the experimental value. The line shows equation 10. The numbers refer to the table in the SI.*

**References**

1. N. Behabtu, J. R. Lomeda, M. J. Green, A. L. Higginbotham, A. Sinitskii, D. V. Kosynkin, D. Tsentalovich, A. N. G. Parra-Vasquez, J. Schmidt, E. Kesselman, Y. Cohen, Y. Talmon, J. M. Tour and M. Pasquali, *Nature nanotechnology*, 2010, **5**, 406-411.
2. V. Chabot, B. Kim, B. Sloper, C. Tzoganakis and A. Yu, *Scientific reports*, 2013, **3**, 1378.
3. J. N. Coleman, *Accounts of chemical research*, 2013, **46**, 14-22.
4. L. Dong, S. Lin, L. Yang, J. Zhang, C. Yang, D. Yang and H. Lu, *Chemical communications (Cambridge, England)*, 2014, **50**, 15936-15939.
5. A. A. Green and M. C. Hersam, *Nano letters*, 2009, **9**, 4031-4036.

6. Y. Hernandez, V. Nicolosi, M. Lotya, F. M. Blighe, Z. Sun, S. De, I. T. McGovern, B. Holland, M. Byrne, Y. K. Gun'Ko, J. J. Boland, P. Niraj, G. Duesberg, S. Krishnamurthy, R. Goodhue, J. Hutchison, V. Scardaci, A. C. Ferrari and J. N. Coleman, *Nature nanotechnology*, 2008, **3**, 563-568.
7. V. Nicolosi, M. Chhowalla, M. G. Kanatzidis, M. S. Strano and J. N. Coleman, *Science*, 2013, **340**, 1226419-1226419.
8. K. R. Paton, E. Varrla, C. Backes, R. J. Smith, U. Khan, A. O'Neill, C. Boland, M. Lotya, O. M. Istrate, P. King, T. Higgins, S. Barwich, P. May, P. Puczkarski, I. Ahmed, M. Moebius, H. Pettersson, E. Long, J. Coelho, S. E. O'Brien, E. K. McGuire, B. M. Sanchez, G. S. Duesberg, N. McEvoy, T. J. Pennycook, C. Downing, A. Crossley, V. Nicolosi and J. N. Coleman, *Nature materials*, 2014, **13**, 624-630.
9. F. Torrisi and J. N. Coleman, *Nature nanotechnology*, 2014, **9**, 738-739.
10. F. Torrisi, T. Hasan, W. Wu, Z. Sun, A. Lombardo, T. S. Kulmala, G.-W. Hsieh, S. Jung, F. Bonaccorso, P. J. Paul, D. Chu and A. C. Ferrari, *ACS nano*, 2012, **6**, 2992-3006.
11. M. a. Bissett, I. A. Kinloch and R. A. W. Dryfe, *ACS Applied Materials & Interfaces*, 2015, 150731121859008.
12. I.-W. P. Chen, Y.-S. Chen, N.-J. Kao, C.-W. Wu, Y.-W. Zhang and H.-T. Li, *Carbon*, 2015, **90**, 16-24.
13. D. Hanlon, C. Backes, T. M. Higgins, M. Hughes, A. O'Neill, P. King, N. McEvoy, G. S. Duesberg, B. Mendoza Sanchez, H. Pettersson, V. Nicolosi and J. N. Coleman, *Chemistry of Materials*, 2014, **26**, 1751-1763.
14. T. M. Higgins, D. McAteer, J. C. M. Coelho, B. M. Sanchez, Z. Gholamvand, G. Moriarty, N. McEvoy, N. C. Berner, G. S. Duesberg, V. Nicolosi and J. N. Coleman, *ACS nano*, 2014, **8**, 9567-9579.
15. B. Mendoza-Sánchez, T. Brousse, C. Ramirez-Castro, V. Nicolosi and P. S. Grant, *Electrochimica Acta*, 2013, **91**, 253-260.
16. Z. Yang, J. Deng, X. Chen, J. Ren and H. Peng, *Angewandte Chemie International Edition*, 2013, **52**, 13453-13457.
17. D. J. Finn, M. Lotya, G. Cunningham, R. J. Smith, D. McCloskey, J. F. Donegan and J. N. Coleman, *J. Mater. Chem. C*, 2014, **2**, 925-932.
18. G. S. Bang, K. W. Nam, J. Y. Kim, J. Shin, J. W. Choi and S. Y. Choi, *ACS Applied Materials and Interfaces*, 2014, **6**, 7084-7089.
19. J. D. Benck, T. R. Hellstern, J. Kibsgaard, P. Chakthranont and T. F. Jaramillo, *ACS Catalysis*, 2014, **4**, 3957-3971.
20. Q. Ding, F. Meng, C. R. English, M. Cabán-Acevedo, M. J. Shearer, D. Liang, A. S. Daniel, R. J. Hamers and S. Jin, *Journal of the American Chemical Society*, 2014, **136**, 8504-8507.
21. A. Harvey, C. Backes, Z. Gholamvand, D. Hanlon, D. McAteer, H. C. Nerl, E. McGuire, A. Seral-Ascaso, Q. M. Ramasse, N. McEvoy, S. Winters, N. C. Berner, D. McCloskey, J. F. Donegan, G. S. Duesberg, V. Nicolosi and J. N. Coleman, *Chemistry of Materials*, 2015, **27**, 3483-3493.
22. M. A. Lukowski, A. S. Daniel, F. Meng, A. Forticaux, L. Li and S. Jin, *Journal of the American Chemical Society*, 2013, **135**, 10274-10277.
23. L. Gong, R. J. Young, I. a. Kinloch, I. Riaz, R. Jalil and K. S. Novoselov, *ACS Nano*, 2012, **6**, 2086-2095.
24. C. Vallés, A. M. Abdelkader, R. J. Young and I. a. Kinloch, *Composites Science and Technology*, 2015, **111**, 17-22.
25. S. Xie, O. M. Istrate, P. May, S. Barwich, A. P. Bell, U. Khan and J. N. Coleman, *Nanoscale*, 2015, **7**, 4443-4450.
26. R. J. Young, I. a. Kinloch, L. Gong and K. S. Novoselov, *Composites Science and Technology*, 2012, **72**, 1459-1476.
27. C. Backes, R. J. Smith, N. McEvoy, N. C. Berner, D. McCloskey, H. C. Nerl, A. O'Neill, P. J. King, T. Higgins, D. Hanlon, N. Scheuschner, J. Maultzsch, L. Houben, G. S. Duesberg, J. F. Donegan, V. Nicolosi and J. N. Coleman, *Nature communications*, 2014, **5**, 4576.
28. M. Lotya, P. J. King, U. Khan, S. De and J. N. Coleman, *ACS Nano*, 2010, **4**, 3155-3162.

29. O. Y. Posudievsky, O. a. Khazieieva, V. V. Cherepanov, V. G. Koshechko and V. D. Pokhodenko, *Journal of Nanoparticle Research*, 2013, **15**, 2046.
30. R. R. Nair, P. Blake, A. N. Grigorenko, K. S. Novoselov, T. J. Booth, T. Stauber, N. M. R. Peres and A. K. Geim, *Science*, 2008, **320**, 1308.
31. C. Backes, K. R. Paton, D. Hanlon, S. Yuan, M. I. Katsnelson, J. Houston, R. J. Smith, D. McCloskey, J. F. Donegan and J. N. Coleman, *Nanoscale*, 2015, **submitted**.
32. Y. Arao and M. Kubouchi, *Carbon*, 2015.
33. M. Ayán-Varela, J. I. Paredes, L. Guardia, S. Villar-Rodil, J. M. Munuera, M. Díaz-González, C. Fernández-Sánchez, A. Martínez-Alonso and J. M. D. Tascón, *ACS Applied Materials & Interfaces*, 2015, 150506130806005.
34. R. Bari, D. Parviz, F. Khabaz, C. D. Klaassen, S. D. Metzler, M. J. Hansen, R. Khare and M. J. Green, *Phys. Chem. Chem. Phys.*, 2015, **17**, 9383-9393.
35. A. Ciesielski, S. Haar, M. El Gemayel, H. Yang, J. Clough, G. Melinte, M. Gobbi, E. Orgiu, M. V. Nardi, G. Ligorio, V. Palermo, N. Koch, O. Ersen, C. Casiraghi and P. Samorì, *Angewandte Chemie (International ed. in English)*, 2014, **53**, 10355-10361.
36. J. N. Coleman, M. Lotya, A. O'Neill, S. D. Bergin, P. J. King, U. Khan, K. Young, A. Gaucher, S. De, R. J. Smith, I. V. Shvets, S. K. Arora, G. Stanton, H.-Y. Kim, K. Lee, G. T. Kim, G. S. Duesberg, T. Hallam, J. J. Boland, J. J. Wang, J. F. Donegan, J. C. Grunlan, G. Moriarty, A. Shmeliov, R. J. Nicholls, J. M. Perkins, E. M. Grieveson, K. Theuwissen, D. W. McComb, P. D. Nellist and V. Nicolosi, *Science (New York, N.Y.)*, 2011, **331**, 568-571.
37. U. Khan, A. O'Neill, M. Lotya, S. De and J. N. Coleman, *Small*, 2010, **6**, 864-871.
38. S. Lin, C.-J. Shih, M. S. Strano and D. Blankschtein, *Journal of the American Chemical Society*, 2011, **133**, 12810-12823.
39. M. Lotya, Y. Hernandez, P. J. King, R. J. Smith, V. Nicolosi, L. S. Karlsson, F. M. Blighe, S. De, W. Zhiming, I. T. McGovern, G. S. Duesberg and J. N. Coleman, *Journal of the American Chemical Society*, 2009, **131**, 3611-3620.
40. P. May, U. Khan, J. M. Hughes and J. N. Coleman, *The Journal of Physical Chemistry C*, 2012, **116**, 11393-11400.
41. D. Nuvoli, L. Valentini, V. Alzari, S. Scognamillo, S. B. Bon, M. Piccinini, J. Illescas and A. Mariani, *Journal of Materials Chemistry*, 2011, **21**, 3428.
42. A. O'Neill, U. Khan and J. N. Coleman, *Chemistry of Materials*, 2012, **24**, 2414-2421.
43. R. J. Smith, P. J. King, M. Lotya, C. Wirtz, U. Khan, S. De, A. O'Neill, G. S. Duesberg, J. C. Grunlan, G. Moriarty, J. Chen, J. Wang, A. I. Minett, V. Nicolosi and J. N. Coleman, *Advanced Materials*, 2011, **23**, 3944 - 3948.
44. L. Xu, J. W. McGraw, F. Gao, M. Grundy, Z. Ye, Z. Gu and J. L. Shepherd, *Journal of Physical Chemistry C*, 2013, **117**, 10730-10742.
45. M. Yi, Z. Shen, X. Zhang and S. Ma, *Journal of Physics D: Applied Physics*, 2013, **46**, 025301.
46. L. Zhang, Z. Zhang, C. He, L. Dai, J. Liu and L. Wang, *ACS nano*, 2014, **8**, 6663-6670.
47. R. Zhang, B. Zhang and S. Sun, *RSC Adv.*, 2015, **5**, 44783-44791.
48. W. Zhang, Y. Wang, D. Zhang, S. Yu, W. Zhu, J. Wang, F. Zheng, S. Wang and J. Wang, *Nanoscale*, 2015, **7**, 10210-10217.
49. H.-L. Liu, C.-C. Shen, S.-H. Su, C.-L. Hsu, M.-Y. Li and L.-J. Li, *Applied Physics Letters*, 2014, **105**, 201905.